%% file: enten_arxiv.tex
\documentclass[prl,twocolumn,lengthcheck]{revtex4-1}

\usepackage{amsmath}
\usepackage{amssymb}
\usepackage{graphicx}
\usepackage[colorlinks,citecolor=magenta,linkcolor=blue]{hyperref}
\usepackage{color}
\usepackage{braket}

\newcommand{\tr}{\operatorname{tr}}

\newlength{\figheight}
\newlength{\figwidth}
\usepackage{tikz,pgfplots}

\begin{document}

\title{Energy cost of entanglement extraction in complex quantum systems}

\author{C\'edric B\'eny$^1$}
\author{Christopher T.\ Chubb$^{2}$}
\author{Terry Farrelly$^{3}$}
\author{Tobias J.\ Osborne$^{3}$}
\affiliation{$^{1}$Department of Applied Mathematics, Hanyang University (ERICA), 55 Hanyangdaehak-ro, Ansan, Gyeonggi-do, 426-791, Korea.\\
$^{2}$Centre for Engineered Quantum Systems, School of Physics, University of Sydney, Sydney, NSW 2006, Australia.\\
$^{3}$Institut f\"ur Theoretische Physik, Leibniz Universit\"at Hannover, Appelstr. 2, 30167 Hannover, Germany.
}

\begin{abstract}
  What is the energy cost of extracting entanglement from complex quantum systems?  In other words, given a state of a quantum system, how much energy does it cost to extract $m$ EPR pairs?  This is an important question, particularly for quantum field theories where the vacuum is generally highly entangled.  Here we build a theory to understand the energy cost of entanglement extraction.  First, we apply it to a toy model, and then we define the \textit{entanglement temperature}, which relates the energy cost to the amount of extracted entanglement.  Next, we give a physical argument to find the energy cost of entanglement extraction in some condensed matter and quantum field systems.  The energy cost for those quantum field theories depends on the spatial dimension, and in one dimension, for example, it grows exponentially with the number of EPR pairs extracted.  Next, we outline some approaches for bounding the energy cost of extracting entanglement in general quantum systems.  Finally, we look at the antiferromagnetic Heisenberg and transverse field Ising models numerically to calculate the entanglement temperature using matrix product states.
\end{abstract}

\maketitle

\section{Introduction}
Entanglement is the most important resource in quantum information, with a vast array of practical uses \cite{HHHH09}.  In physics more generally, understanding the entanglement structure in physical systems is becoming increasingly important, e.g., in condensed matter theory where quantum phase transitions are signalled by long-range entanglement \cite{OAFF01,ON02,VLRK03,AFOV08,LR09}.  In high energy physics, quantifying the entanglement of states of a quantum field has applications to a variety of problems \cite{SW85,Summers08,KST17}, from the AdS/CFT correspondence \cite{RT06}, through to detecting spacetime curvature by probing vacuum entanglement \cite{SM09} or harvesting this entanglement by locally coupling small systems to the field \cite{Valentini91,Reznik03,RRS05,SR07,PM15,PM16,SM17,SMM17}.

Quantifying entanglement in states of quantum fields is a nontrivial task due to the UV dependence of quantum entanglement near a boundary, leading to naive divergences \cite{HLW94,CC04,CC06}.  Many ad hoc approaches have been developed to deal with these divergences, usually relying on subtracting the UV divergent piece \cite{HLW94}.  Of course, operationally, there are no such divergences in the entanglement we can measure because any apparatus we can build to extract entanglement from the field vacuum would only use a finite amount of energy.

Surprisingly little work has been done in the quantum information literature on the problem of quantifying accessible entanglement subject to an energy constraint.  Here we build a theory for this problem and use it to understand the energy cost of extracting entanglement via local operations and classical communication (LOCC).  We use the term extraction rather than one-shot distillation \cite{BD10}, as we want to emphasise that we are not necessarily distilling all the entanglement from a state.  In contrast, we wish to quantify the \textit{optimal} energy cost per EPR pair extracted.  While individual protocols for entanglement extraction are interesting, we are primarily concerned with the protocol that minimises the energy cost.

There are some very interesting related ideas in the literature:\ in \cite{CY17}, general quantum operations costing \textit{zero} energy are studied.  Also, the energy cost of \textit{creating} entanglement in specific many-body systems was calculated in \cite{GL09}.  Similarly, in the setting of quantum thermodynamics, the energy/work cost of creating correlations in quantum systems was studied in \cite{HPHSKBA15,BPFHH15,FHP16}.  These give useful strategies for creating correlations between finite dimensional systems or a pair of bosons or fermions using energy-conserving (global) unitary operations in the presence of heat baths.  In \cite{DKPSSS17}, entanglement distillation is considered (also in the presence of a heat bath) with an energy constraint:\ asymptotically many entangled pairs are distilled into EPR pairs, with the constraint that the energy before and after is \textit{equal}.  In \cite{MBDK15}, using a specific local entanglement harvesting protocol (called entanglement farming), the energy cost in the low energy regime was calculated.  In contrast, here we are interested in how the \textit{optimal} energy cost scales with the number of EPR pairs extracted and in the overall entanglement structure of states, which is a rather different question.

In this article, we first provide the setting and define the energy cost of a quantum operation.  Then we introduce entanglement extraction subject to an energy constraint.  Following this, we explore the idea using a toy model, which is chosen to share many of the features of the vacuum state of a quantum field but to also be relatively simple.  Next, we introduce the entanglement temperature, which relates the amount of entanglement extracted to the energy cost.  In the following section, we look at the energy cost of entanglement extraction in quantum field theories using physical arguments.  Then we discuss some general methods for quantifying the energy cost of entanglement extraction.  In particular, one of these methods uses matrix product states to numerically bound the energy cost of entanglement extraction for some condensed matter systems and to plot the entanglement temperature.  We conclude with an outlook.

\section{Preliminaries}
We focus on a simplified setup exemplifying the core features of our problem. To whit, we focus on the setting where two players, Alice and Bob, have access to a bipartition of a common system with Hilbert space $\mathcal{H}_{AB}$. This system, which we refer to as the \textit{physical system}, has a Hamiltonian $H_{AB}$, which neither Alice nor Bob can modify. Alice and Bob also have access to local ancillary degrees of freedom $A'B'$, which they can use to store the entanglement they extract from the physical system. Thus, the total Hilbert space for the system is given by 
\begin{equation}
	\mathcal{H}_{AA'BB'} \equiv \mathcal{H}_{AB}\otimes \mathcal{H}_{A'B'}. 
\end{equation} 
We assume that Alice and Bob can carry out any local operation they like on the ancillary degrees of freedom with no energy cost. Thus we associate to the total system+ancilla system the Hamiltonian $H=H_{AB}\otimes\openone_{A'B'}$.
\begin{figure}[ht!]
 \resizebox{8.5cm}{!}{\includegraphics{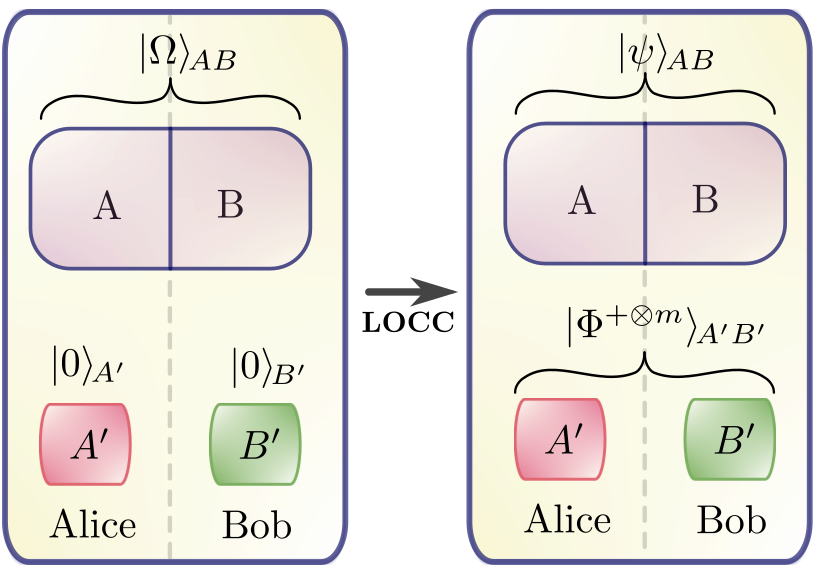}}
    \footnotesize{\caption[Basic Idea]{Energy cost of entanglement extraction:\ Alice and Bob have access to two parts of a quantum system in an entangled state $\ket{\Omega}_{AB}$ with Hamiltonian $H_{AB}$.  Using local operations and classical communication (LOCC), they extract $m$ EPR pairs into their ancillary systems $A^{\prime}$ and $B^{\prime}$, leaving the physical system in the final state $\ket{\psi}_{AB}$.  Dropping subscripts, the energy cost is then $\Delta E = \bra{\psi}H\ket{\psi}-\bra{\Omega}H\ket{\Omega}$. \label{fig:intro}}}
\end{figure}

We assume that Alice and Bob can only perform local operations and classical communication (LOCC).  We also suppose that Alice and Bob are working in the one-shot regime, which is natural if, for example, we are thinking of understanding the entanglement structure of vacuum states in quantum field theory where there is only one copy of the system available.  In contrast, in the asymptotic many-copy regime we could use entanglement distillation protocols \cite{HHHH09}.  We will also comment on the energy cost in the asymptotic regime.

We quantify the energy cost as follows.  Suppose we have a completely positive trace-preserving map $\mathcal{E}:\mathcal{S}(\mathcal{H})\rightarrow \mathcal{S}(\mathcal{H})$ acting on the space $\mathcal{S}(\mathcal{H})$ of density operators on a Hilbert space $\mathcal{H}$. Suppose we also have a Hamiltonian $H \in \mathcal{B}(\mathcal{H})$. Given a state $\rho \in \mathcal{S}(\mathcal{H})$, the operation $\mathcal{E}$ induces an energy change 
\begin{equation}
	\Delta(\mathcal{E}, \rho) = \tr(\mathcal{E}(\rho)H) - \tr(\rho H)
\end{equation}
when acting on $\rho$ (this can be negative).  This is the energy cost when we apply the channel $\mathcal{E}$ to the state $\rho$.  We may also define the \textit{energy cost for the channel} $\mathcal{E}$, which involves an optimisation:\ we imagine that an adversary prepares the system in the state $\rho$ which leads to the \textit{largest} possible change in energy after application of $\mathcal{E}$:
\begin{equation}
	\Delta(\mathcal{E}) = \sup_{\rho \in \mathcal{S}(\mathcal{H})} \tr(\mathcal{E}(\rho)H) - \tr(\rho H).
\end{equation}
Exploiting the variational definition of the operator norm $\|\cdot\|_\infty$, we notice that
\begin{equation}
	\Delta(\mathcal{E}) = \|H-\mathcal{E}^*(H)\|_\infty,
\end{equation}
where $\mathcal{E}^*$ is the dual of $\mathcal{E}$ acting in the Heisenberg picture.  In the following, we will typically be interested in $\Delta(\mathcal{E},\rho)$ for a specific state and channel, or the set of LOCC channels with $\Delta(\mathcal{E},\rho)\leq \Delta E$.  We denote this set by $\mathcal{C}_{\text{LOCC}}(\Delta E)$.

\section{Extracting entanglement subject to an energy constraint}
Here we propose a definition for the entanglement accessible to Alice and Bob when they have access only to operations costing less than $\Delta E$. 

We imagine that the system $AB$ starts in a state $\sigma_{ABA^{\prime}B^{\prime}}=|\Omega\rangle_{AB}\bra{\Omega}\otimes|00\rangle_{A'B'}\!\bra{00}$ where $|0\rangle$ is a convenient fiducial state of the ancilla and $|\Omega\rangle_{AB}$ is the initial state of the physical system. Alice and Bob are now allowed to carry out LOCC operations costing less than $\Delta E$ in total to maximise the quantum entanglement between $A'$ and $B'$. Suppose that $\mathcal{E}\in \mathcal{C}_{\text{LOCC}}(\Delta E)$, then we write $\rho_{AA'BB'} = \mathcal{E}(\sigma_{ABA^{\prime}B^{\prime}})$.  Thus we define the \emph{entanglement accessible with energy $\Delta E$} to be
\begin{equation}
	\text{Ent}_{\Delta E}(|\Omega_{AB}\rangle) \equiv \sup_{\mathcal{E}\in \mathcal{C}_{\text{LOCC}}(\Delta E)} \text{Ent} (\rho_{A'B'}),
\end{equation}
where $\text{Ent}$ is some convenient entanglement measure.

We also define the \emph{energy cost of extracting $m$ EPR pairs} to be $\Delta E=\min\Delta\left(\mathcal{E},\sigma_{ABA^{\prime}B^{\prime}}\right)$, where the minimum is over all LOCC channels satisfying $\rho_{A'B'}=\ket{\phi^{+\otimes m}}_{A^{\prime}B^{\prime}}\bra{\phi^{+\otimes m}}$, and $\ket{\phi^{+}}=\frac{1}{\sqrt{2}}(|00\rangle + |11\rangle)$ is a maximally entangled state of two qubits.

It is well possible that after extracting entanglement the energy of the system can \textit{go down}, i.e., extracting entanglement can cool the system. This all depends on the state $|\Omega_{AB}\rangle$, i.e., whether it is an excited state or ground state. Since the emphasis in this paper is on ground states, we assume henceforth that $|\Omega_{AB}\rangle$ is the ground state of $H_{AB}$.

A key ingredient in any entanglement extraction protocol is the strength of the interaction between Alice's and Bob's systems.  If we write $H_{AB}=H_A\otimes \openone_B + \openone_A \otimes H_B + V_{AB}$, then the limitations on how much entanglement Alice and Bob can extract using LOCC are determined by $V_{AB}$.  Indeed, if $V_{AB}=0$, the ground state $\ket{\Omega}$ will have no entanglement between Alice's and Bob's systems.

There is a useful naive protocol for entanglement extraction:\ Alice and Bob first swap the states of their primed and non-primed systems.  Then they can prepare a state of the physical $AB$ system (using LOCC) with minimal \textit{local} energy, meaning Alice/Bob prepares $\ket{\psi_{A/B}}$, such that $\bra{\psi_{A/B}} H_{A/B} \ket{\psi_{A/B}}$ is minimised.  The total energy change is, with $\ket{\psi}=\ket{\psi_A}\ket{\psi_B}$,
\begin{equation}
\begin{split}
 & \bra{\psi} H \ket{\psi}- \bra{\Omega} H\ket{\Omega}\\
 \leq & \bra{\psi} V_{AB} \ket{\psi}- \bra{\Omega} V_{AB}\ket{\Omega} \leq 2\|V_{AB}\|_{\infty}.
 \end{split}
\end{equation}
Therefore, when the coupling is sufficiently weak, Alice and Bob can safely extract all the entanglement whilst only incurring a small energy cost.

In contrast, for strong couplings the situation is entirely different, which is exactly the case for quantum field theories, where extracting all the entanglement costs a divergent amount of energy.  For the example of a free fermion field, we see in the appendix that all product states $\ket{\psi}$ satisfy $\bra{\psi}H\ket{\psi}\geq 1/a$, where $a$ is the regulator (the lattice spacing in this case).  Thus, the energy diverges as $a\rightarrow 0$ for any product state, meaning that extracting all the entanglement costs a diverging amount of energy.  In general, the energy cost for extracting all the entanglement will diverge in quantum field theory.  Again using a lattice regulator, the energy contained in the interaction terms between a region $A$ and the rest scales like $(\partial A/a^{d})$, where $\partial A$ is the boundary of $A$, which is also shown in the appendix.

\section{A toy model}
In this section we discuss an idealised model, which exemplifies many of the features of the quantum field vacuum.  It has high entanglement and a high energy cost for extracting all this entanglement, as we will see. 

Suppose that the system $AB$ is actually composed of $2n$ qubits, with $n$ qubits in $A$ and $n$ qubits in $B$. We call the qubits $A_j$ (respectively, $B_j$), for $j=1,2, \ldots, n$. We suppose that $H_{AB}$ is given by 
\begin{equation}
	H_{AB} = \sum_{j=1}^n (\mathbb{I}-P_{A_jB_j}),
\end{equation}
where $P_{A_jB_j}$ is the projector onto the maximally entangled state $\ket{\Phi^{+}}=\frac{1}{\sqrt{2}}(|00\rangle + |11\rangle)$ of qubits $A_j$ and $B_j$. The ground state $|\Omega_{AB}\rangle$ of $H_{AB}$ is thus a product of maximally entangled pairs, i.e., it is a maximally entangled state between $A$ and $B$.

If Alice and Bob could do arbitrary LOCC, then they could easily extract $n$ EPR pairs. However, if they are only allowed an energy cost of $\Delta E$, then naively they should only be able to extract $O(\Delta E)$ EPR pairs. 

In the most extreme case, Alice and Bob fully extract all the EPR pairs. Then, in order that this entanglement is between ancilla degrees of freedom in $A'$ and $B'$, it must be that $A$ and $B$ are in a separable state $\sigma_{AB}$. Since the energy depends linearly on $\sigma_{AB}$, we may as well suppose that $\sigma_{AB}$ is an extreme point of the convex set of separable states, namely, a product state $|\phi\rangle_A|\psi\rangle_B$. The energy of our initial state $|\Omega\rangle_{AB}$ was zero, so the energy cost of any entanglement extraction procedure must be greater than  
\begin{equation}
	\inf_{|\phi\rangle_A|\psi\rangle_B} \sum_{j=1}^n (1-\langle \phi_A|\langle \psi_B|P_{A_jB_j}|\phi_A\rangle|\psi_B\rangle).
\end{equation}
This infimum is achieved by finding the supremum:
\begin{equation}
 	\sup_{|\phi\rangle_A|\psi\rangle_B} \langle \phi_A|\langle \psi_B|P_{A_jB_j}|\phi_A\rangle|\psi_B\rangle,
\end{equation} 
which is equal to $1/2$.  (E.g., setting each pair to $\ket{00}_{A_iB_i}$ will do the job.) Thus, the energy cost is given by
\begin{equation}
	\Delta E \ge \frac{1}{2}\sum_{j=1}^n 1 = n/2.
\end{equation}

More generally, suppose Alice and Bob extract fewer EPR pairs (say $m$).  One option is to use the following simple protocol.  They swap the states of the first $m$ EPR pairs of the physical system into their ancilla systems, which they can do using local operations.  The first $m$ pairs of qubits of the physical system are now in a product pure state.  Then they can apply local unitaries mapping each of these qubit pairs to the state $\ket{00}_{A_iB_i}$, getting the final energy cost
\begin{equation}
	\Delta E = \frac{1}{2}\sum_{j=1}^m 1 = m/2.
\end{equation}
Thus, the total energy cost is $1/2$ per EPR pair extracted.

Of course, there may be a protocol extracting the same amount of entanglement but costing less energy.  Here we will argue that the simple protocol given above is, in fact, optimal.

We assume that after applying their operations, Alice and Bob get $m$ Bell states in the ancilla $\ket{\Phi^{+\otimes m}}_{A'B'}$ and some pure state in the physical system $\ket{\psi}_{AB}$.  Denote the Schmidt values (decreasingly ordered) of the initial state $\ket{\Omega}_{AB}\ket{00}_{A'B'}$ by $\alpha_i$, and note that the Schmidt rank is $2^n$.  For it to be possible to transform this state into the new state $\ket{\psi}_{AB}\ket{\Phi^{+\otimes m}}_{A'B'}$, with Schmidt values $\beta_i$, the majorization condition \cite{NC00} must be satisfied:
\begin{equation}
 \forall K\geq 1\ \ \ \  \sum_{i=1}^{K}\alpha_i\leq \sum_{i=1}^{K}\beta_i.
\end{equation}
For this to be possible, the Schmidt rank of the resulting state must be smaller.  This implies that the Schmidt rank of the new state of the physical system $\ket{\psi}_{AB}$ can be at most $2^{n-m}$.

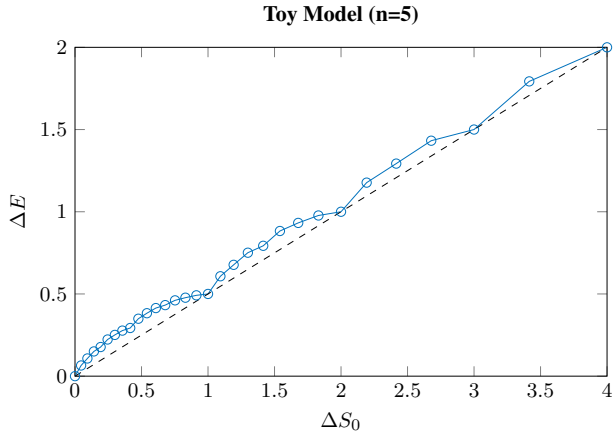
\begin{figure}[ht!]
 \setlength{\figheight}{5cm}
 \setlength{\figwidth}{8.5cm}
 \def\figscale{.875}
 \input{./Toy.tikz}  
    \footnotesize{\caption[Toy Model]{This figure shows the minimum energy cost of extracting entanglement for the toy model.  More precisely, the figure shows the minimum change in energy $\Delta E$ when there is a decrease in zero entropy $\Delta S_0$ (with $S_0$ equal to $\log_2$ of the Schmidt rank) of the state of the physical system.\label{fig:toymodel}  When $\Delta S_0$ is an integer $m$ (which corresponds to extracting $m$ EPR pairs from the system), then the plot shows $\Delta E =0.5m$.  The calculation was performed using DMRG by restricting the bond dimension between Alice and Bob's systems.}}
\end{figure}

Figure \ref{fig:toymodel} shows numerics from a DMRG calculation of the minimum increase in energy as the Schmidt rank of the state of the physical system decreases.  Based on these numerical results, we see that the minimum increase in energy when the Schmidt rank decreases by a factor of $2^m$ is $0.5 m$.  This can be achieved by the simple protocol of the previous section, indicating that this protocol is optimal.  Actually, this whole argument also goes through even if Alice and Bob have some additional shared entanglement that can be used as a catalyst, as in \cite{JP99}.

In terms of entanglement distillation in the asymptotic setting, this is not optimal.  In that case, one can distil entanglement at a lower energy cost, which we show later.  In practice, however, we only have access to one copy of a quantum field or condensed matter system, so it is crucial to consider the one-shot setting.  Furthermore, entanglement distillation protocols rely on projecting onto typical subspaces defined by the singular vectors of the initial state \cite{NC00}, which for extremely complex systems would be practically impossible.

\section{The entanglement temperature}
In the previous section, the total energy cost was $1/2$ per EPR pair extracted.  To relate the change in entanglement entropy $\Delta S$ to the energy cost $\Delta E$, we define the \textit{entanglement temperature} $T_{\mathrm{ent}}$ by
\begin{equation}
 \Delta E = T_{\mathrm{ent}} \Delta S.
\end{equation}
(The name entanglement temperature is chosen in analogy with thermodynamics.)  So $T_{\mathrm{ent}}$ is a property of the ground state of a system.  For the toy model, we see that $T_{\mathrm{ent}}=1/2$ since $\Delta S=m\log_{2}(2)=m$.  In this case $T_{\mathrm{ent}}$ is constant because there is a linear relationship between the entanglement extracted and the energy cost.  For general systems, we would not expect $\Delta E \propto \Delta S$ for the entire range of $\Delta S$.  Instead, we should think of the entanglement temperature as a function of the extracted entanglement.  (This is also true in thermodynamics, where temperature can often be thought of as a function of other state functions, such as entropy or pressure.)

In the following sections, we give some physical and numerical arguments to find $\Delta E$ as a function of $\Delta S$ and hence find the entanglement temperature for some physical systems.

\section{The Energy Cost in General}
\label{sec:General Strategy}
For some quantum field theories or condensed matter systems, we can give a physical argument for the energy cost of entanglement extraction.  In one dimensional systems, often the entanglement entropy (or, for example, the logarithmic negativity) of ground states can be calculated.  This typically has the form $S(\rho_I)=c_1\log_2(N)+c_2$, where $c_i$ are constants and $\rho_I$ is the state restricted to a contiguous region with $N$ sites \cite{Korepin04,ECP10}.  For a quantum field theory, regulated by a lattice with lattice spacing $a$, we have instead $S(\rho_I)=c_1\log_2(l/a)+c_2$, where $l$ is the length of a region.  Also, the entanglement entropy in the ground state of models close to the critical point is \cite{CC04,ECP10}
\begin{equation}
 S(\rho_I)=\frac{c}{6}\log_2(\xi/a),
\end{equation}
where $\xi\gg a$ is the correlation length, $c$ is a constant and $I$ corresponds to the infinite half-line $(-\infty,0]$.  This is equivalent to a massive relativistic QFT with $1/\xi$ equal to the mass, e.g., for free bosons we have $c=1$.

As argued previously, the energy cost for extracting all of this entanglement is $\Delta E = O(1/a)$.  At lattice spacing, $a=\xi/2^{6m/c}$, we have $S(\rho_I)=m$, meaning that the \textit{most} entanglement we can extract is $m$ EPR pairs.  (In some cases, exactly half of this entanglement is one-shot distillable \cite{OLEC06}.)  By probing higher energies, which corresponds to smaller values of $a$, we can extract more entanglement, and we have the energy cost $\Delta E\propto \exp(Km)$, where $K=6\ln(2)/c$.  This means that the energy cost of entanglement extraction increases \textit{exponentially}:\ there is infinite entanglement in the quantum field vacuum, but the cost of extraction grows quickly.  (For gapless models the same argument goes through if Alice has access to a finite region and Bob has access to the rest.  In contrast, if Alice's system is a halfline, there is infinite entanglement at any energy scale.)

\begin{figure}
	\setlength{\figheight}{4cm}
	\setlength{\figwidth}{8cm}
	\def\figscale{1}
	\!\!\input{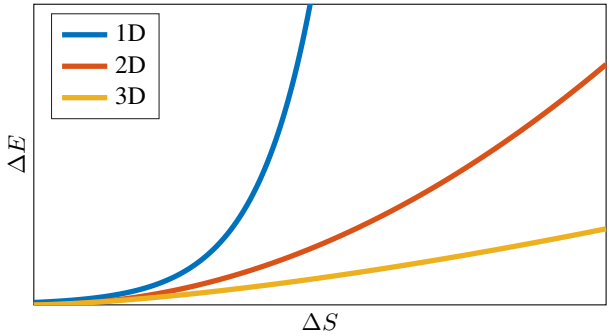}
	\footnotesize{\caption{The energy cost of entanglement extraction from the quantum field vacuum depends heavily on the spatial dimension.  Here we sketch the behaviour in dimensions $d=1,2,3$.  When $d=1$, $\Delta E\propto \exp(K\Delta S)$, where $K$ is a constant, and for $d>1$, $\Delta E \propto \Delta S^{d/(d-1)}$.}\label{fig:qft}}
\end{figure}

The scaling is different for quantum fields in higher dimensional spaces.  In many cases, the entanglement entropy of the ground state obeys an area law \cite{RT06b}.  Then in a region $A$ with area $\partial A$, the leading contribution to the entropy is $S(\rho_A)\propto \partial A/a^{d-1}$.  However, the energy cost of extracting all the entanglement scales like $\Delta E \propto \partial A/a^{d}$.  Thus we get an idea for the energy cost of entanglement extraction:\ $\Delta E \propto \Delta S^{d/(d-1)}$.  And the entanglement temperature is then $T_{\mathrm{ent}}\propto 1/a\propto \Delta E^{1/d}$.  The energy cost of entanglement extraction in QFT are plotted in figure \ref{fig:qft}.

For more general systems, there is no clear way to proceed.  Below we outline some potential methods to approach the problem.  In the first two cases, we suppose that Alice and Bob use some LOCC protocol to extract $m$ EPR pairs into their ancillary systems, and in the third we consider the trade-off between entanglement change and energy cost numerically.

\subsection{Method I}
Alice and Bob have the initial state $|\Omega\rangle_{AB}\ket{00}_{A^{\prime}B^{\prime}}$, and then they apply some LOCC protocol to extract $m$ EPR pairs into the ancilla $A^{\prime}B^{\prime}$.  We assume that the resulting state on the physical system after the protocol is also pure $\ket{\psi}_{AB}$.  (It is possible that a protocol giving a mixed state on the physical system may be more efficient.  In this case, we may use a superadditive entanglement measure, like the squashed entanglement \cite{CW04}, to upper bound the entanglement left in the physical system.)  Because they are using LOCC, the overall entanglement can only decrease:
\begin{equation}
\begin{split}
 \text{Ent}\left(|\psi\rangle_{AB}\ket{\Phi^+}^{\otimes m}_{A^{\prime}B^{\prime}}\right) &\leq\text{Ent}\left(|\Omega\rangle_{AB}\ket{00}_{A^{\prime}B^{\prime}}\right)\\
 & = S_{\mathrm{initial}},
 \end{split}
\end{equation}
where $S_{\mathrm{initial}}$ is the initial entanglement entropy in the state $|\Omega\rangle_{AB}$ and $\text{Ent}$ is an entanglement measure, which we take to be the entanglement entropy, since the states are all pure.  Then we have that the entanglement entropy in the final state of physical system is $\text{Ent}\left(|\psi\rangle_{AB}\right)\leq S_{\mathrm{initial}}-m$.

So what is the minimum energy cost of extracting this entanglement?  We can get an idea by finding the state (or set of states) that have this final value of entanglement entropy while minimising the energy.  In the appendix, we derive the equation
\begin{equation}\label{eq:La}
 \left[H - \mu_1\openone_A\otimes\log(\rho_B) - \mu_1 +\mu_2\right]\ket{\psi}_{AB}=0,
\end{equation}
where $\mu_i$ are Lagrange multipliers and $\mathrm{tr}_A[\ket{\psi}\bra{\psi}]=\rho_B$.

This is difficult to solve in general but may be simplified if we know something about the structure of $H$.  This is the case for the toy model, where $H$ is a sum of commuting terms acting on different pairs of qubits $A_iB_i$.

With the ansatz $\ket{\psi}_{AB}=\ket{\psi_1}_{A_1B_1}\otimes...\otimes \ket{\psi_n}_{A_nB_n}$, we see from equation (\ref{eq:La}) that each $\ket{\psi_i}_{A_iB_i}$ should have the same Schmidt vectors as $\ket{\Phi^{+}}_{A_iB_i}$.  One possible solution is to take all qubit pairs to be in the same state:\ $\ket{\psi}_{AB}=\ket{\phi}_{A_1B_1}\otimes...\otimes \ket{\phi}_{A_nB_n}$, where $\ket{\phi}=\alpha\ket{00}+\beta\ket{11}$.  Then, since $S_{\mathrm{initial}}=n$, one need only solve
\begin{equation}\label{eq:ent}
 n-m= -n[\alpha^2\log_2(\alpha^2)+\beta^2\log_2(\beta^2)]
\end{equation}
for $\alpha$ and $\beta$.  And the corresponding energy cost is $\Delta E = n[1- (\alpha+\beta)^2/2]$.

For example, with $m=n/2$, one gets $\Delta E\simeq 0.38 m$.  This is smaller than the optimal energy cost in the one-shot setting:\ $\Delta E= 0.5 m$.  However, in the one-shot setting, Alice and Bob cannot prepare the state $\ket{\psi}_{AB}$ after extracting $m$ EPR pairs (because $\ket{\psi}_{AB}$ has maximal Schmidt rank).  Interestingly, however, we get a nontrivial \textit{upper} bound on the optimal energy cost of extracting entanglement in the asymptotic setting of many copies of this system.  In the asymptotic setting, the criterion for deciding whether one bipartite entangled pure state can be transformed into another reversibly using LOCC is that the entanglement entropies are the same \cite{NC00}.  So we see that the energy cost of distilling $m$ EPR pairs (per copy of the physical system) in the asymptotic setting will be lower than in the one shot case.

\subsection{Method II}
A second option is to maximise the overlap of the final state of the physical system (after the entanglement has been extracted) with its ground state.  This gives a naive strategy at least.  And for Hamiltonians of the form $H_{AB}=-\ket{\Omega}\bra{\Omega}$, we get an exact answer for the optimal energy cost.

As an example, take $\ket{\Omega}=(1/\sqrt{d})\sum_{i=1}^{d}\ket{i}_A\ket{i}_B$, where $d=2^n$.  Suppose that Alice and Bob extract $m$ EPR pairs using LOCC, leaving a pure state $\ket{\psi}$ in the physical system.  Using the majorization criterion, this can be any state with Schmidt rank up to $K=2^{n-m}$.  To minimise the energy cost, we need to find such a state having maximal overlap with $\ket{\Omega}$.

We may write the optimal $\ket{\psi}$ in its Schmidt basis as $\sum_{i=1}^{K}\alpha_i\ket{a_i}_A\ket{b_i}_B$.  Next notice that 
\begin{equation}
\left(\sum_{i=1}^{d}\bra{i}_A\bra{i}_B\right)\ket{a_i}_A\ket{b_i}_B\leq 1.
\end{equation}
Then we have
\begin{equation}
\langle\Omega|\psi\rangle\leq \frac{1}{\sqrt{d}}\sum_{i=1}^K\alpha_i\leq \sqrt{\frac{K}{d}}=2^{-m/2}.
\end{equation}
Therefore, we see that the energy cost for extracting $m$ EPR pairs is $\Delta E = 1-1/2^m$.

\subsection{Method III}
A third option is to consider the trade-off between entanglement and energy numerically. For a given Hamiltonian $H$, we consider a procedure in which the system starts in the ground state $\ket{\Omega}$, some entanglement is extracted, and the system is left in a final state $\ket{\psi}$. The energy cost of this procedure is \mbox{$\left\langle \psi \middle|H\middle|\psi \right\rangle-\left\langle \Omega \middle|H\middle|\Omega \right\rangle$}, and the extracted entropy is upper bounded by \mbox{$\text{Ent}(\ket{\Omega})-\text{Ent}(\ket{\psi})$}.  In the asymptotic many-copy case, this is exactly the extracted entanglement entropy.  We denote the entanglement temperature in that case by $T_{\mathrm{ent}}^A$.   We have that
\begin{align}
	T^{A}_{\mathrm{ent}}\leq \frac{\left\langle \psi \middle|H\middle|\psi \right\rangle-\left\langle \Omega \middle|H\middle|\Omega \right\rangle}{\text{Ent}(\ket{\Omega})-\text{Ent}(\ket{\psi})}.
\end{align}
Note that for a given amount of extracted entanglement $\Delta S$, the one-shot entanglement temperature is lower bounded by the asymptotic-setting entanglement temperature $T_{\mathrm{ent}}^A \leq T_{\mathrm{ent}}$.

A given state does not necessarily give a tight bound on $T_{\mathrm{ent}}^A$. For this we need to study the optimal trade-off between entanglement and energy, which is given by a \emph{Pareto front}. By randomly generating states with low energy \emph{and} low entanglement, we can numerically evaluate the above upper bound, and use this to compute the Pareto front. We describe a tensor network method for generating such samples in the appendix. In Figure~\ref{fig:pareto} we present numerical results for two 1D spin models.

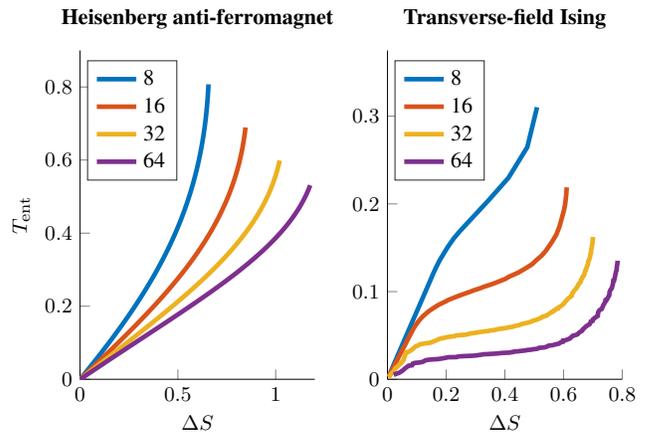
\begin{figure}
	\setlength{\figheight}{5cm}
	\setlength{\figwidth}{3.75cm}
	\def\figscale{.875}
	\! \input{./HAF.tikz} \!\! \input{./TIM.tikz} \!
	\footnotesize{\caption{Asymptotic-case entanglement temperature ($T^A_{\mathrm{ent}}= \Delta E/\Delta S$) as a function of entropy change $\Delta S$ for the critical Heisenberg anti-ferromagnet and the critical transverse-field Ising model, for multiple system sizes. Notice that near the ground state $T^A_{\mathrm{ent}}\propto \Delta S$, which we prove is generic in the appendix.  The one-shot entanglement temperature $T_{\mathrm{ent}}$ is lower bounded by the asymptotic  temperature $T_{\mathrm{ent}}^A$.}\label{fig:pareto}}
\end{figure}

\section{Outlook}
We introduced a framework to understand and quantify the energy cost of extracting entanglement from complex quantum systems.  After looking at a toy model, which illustrated the key concepts, we defined the entanglement temperature.  Then we analysed the energy cost of entanglement extraction in quantum field theories, and we saw that the energy cost of extracting entanglement depends on the spatial dimension.  Finally, we looked at some general methods to approach the problem, including numerical methods for lattice models.  

Quantifying how much energy extracting $m$ EPR pairs costs in physical systems illuminates the entanglement structure of states, particularly ground states of, e.g., quantum fields.  But it also can upper bound how efficient protocols such as entanglement harvesting can be.  For general systems the optimal strategy for entanglement extraction may be hard to find.  Still, it is heartening that, at least for quantum field theories, there is a relatively simple form of the entanglement temperature.

It would be interesting to combine the ideas here with those in \cite{AMOP17}, where transformations between entangled states are considered using an additional resource:\ an entanglement battery.  This is a reservoir from which entanglement may be taken or deposited to facilitate state transformations, which may be impossible otherwise.  One may then ask how this theory changes when there is also an energy cost associated with using the entanglement in the battery.

\section*{Acknowledgements}
We would like to thank David Reeb and Robin Harper for useful discussions.

TF and TJO are supported by the DFG through SFB 1227 (DQ-mat) and the RTG 1991, the ERC grants QFTCMPS and SIQS, and the cluster of excellence EXC201 Quantum Engineering and Space-Time Research. CC acknowledges support from the ARC via the Centre of Excellence in Engineered Quantum Systems (EQuS), project number CE110001013, and from the AINST Postgraduate Scholarship (John Makepeace Bennett Gift).  CB was supported by the research fund of Hanyang University (HY-2016-2237).

\bibliographystyle{unsrt}
\bibliography{enten_arxiv}

\appendix
\section{Energy cost of fermion ground state entanglement}
Consider free massless fermions in $(1+1)$ dimensions with lattice cutoff $a$.  To avoid worrying about fermion doubling, take staggered fermions \cite{Susskind77}, with Hamiltonian
\begin{equation}
 H=\sum_{n=0}^{N-1}\left[\frac{i(\psi^{\dagger}_{n}\psi_{n+1}-\psi^{\dagger}_{n+1}\psi_{n})}{2a}+\frac{1}{a}\right],
\end{equation}
where $\psi_n$ are fermion annihilation operators, satisfying $\{\psi_n,\psi^{\dagger}_{m}\}=\delta_{n,m}$ and $\{\psi_n,\psi_{m}\}=0$.  We have chosen the $1/a$ term on the right so that, as $a\rightarrow 0$, the Hamiltonian is positive definite with ground state energy independent of $a$ (which we verify at the end of the section).  This normalization also has the advantage that each term in the sum is positive definite:
\begin{equation}
\begin{split}
 & i(\psi^{\dagger}_{n}\psi_{n+1}-\psi^{\dagger}_{n+1}\psi_{n})+2\\
 = & (\psi^{\dagger}_{n}-i\psi^{\dagger}_{n+1})(\psi_{n}+i\psi_{n+1})+\psi^{\dagger}_{n}\psi_{n}+\psi^{\dagger}_{n+1}\psi_{n+1}.
 \end{split}
\end{equation}
Now suppose Alice's system is one half of the chain (sites $0$ to $N/2-1$), and Bob's system is the other half.  If they extract the entanglement by swapping the state of the physical system into an ancilla, then the final state of the chain is \footnote{The state can be chosen to be pure because the energy can always be minimised by a pure state.}
\begin{equation}
 \ket{\psi}=(\alpha\psi^{\dagger}_{N/2-1}A_1^{\dagger}+\beta A_2^{\dagger})(\gamma\psi^{\dagger}_{N/2}B_1^{\dagger}+\delta B_2^{\dagger})\ket{0},
\end{equation}
where $\ket{0}$ is the state satisfying $\psi_n\ket{0}=0$ for all $n$; $A_1,A_2$ are products of annihilation operators on Alice's system, while $B_1,B_2$ are products of annihilation operators on Bob's system; and $\alpha,\beta,\gamma,\delta$ are complex numbers.  Because of superselection rules, if $A_1$ is a product of an odd number of annihilation operators, then $A_2$ is even, or vice versa.  The same holds for $B_1$ and $B_2$.  Then one can easily verify that
\begin{equation}
 \bra{\psi}\left[ \frac{i(\psi^{\dagger}_{n}\psi_{n+1}-\psi^{\dagger}_{n+1}\psi_{n})}{2a}+\frac{1}{a}\right]\ket{\psi}\geq \frac{1}{a}.
\end{equation}
So the energy of this state diverges as $a\rightarrow 0$.

Finally, we need to verify that we normalised the Hamiltonian suitably and that the ground state energy of the staggered-fermion Hamiltonian is independent of $a$.  Thus, we need to diagonalise
\begin{equation}
 H=\sum_{n=0}^{N-1}\left[\frac{i(\psi^{\dagger}_{n}\psi_{n+1}-\psi^{\dagger}_{n+1}\psi_{n})}{2a}+\frac{1}{a}\right].
\end{equation}
To do this, we switch to momentum space, with
\begin{equation}
 \psi_n=\frac{1}{\sqrt{N}}\sum_{k=0}^{N-1}e^{-2\pi i k n/N}\psi_k.
\end{equation}
Then we have
\begin{equation}
 H=\sum_{k=0}^{N-1}\left[-\frac{\sin(2\pi k/N)}{a}\psi^{\dagger}_{k}\psi_{k}+\frac{1}{a}\right].
\end{equation}
This has minimum energy
\begin{equation}
E_0=\frac{N}{a} -\frac{N}{2a}\left[\frac{2}{N}\sum_{k=0}^{N/2-1}\sin(2\pi k/N)\right],
\end{equation}
where we have assumed that $N$ is even.  We next define the Riemann sum over $[0,1]$ to be
\begin{equation}\label{eq:Riemann}
 R_M[f(x)]=\frac{1}{M}\sum_{m=1}^{M}f\left(\frac{m-1/2}{M}\right).
\end{equation}
And we may use the following formula for convergence of a Riemann sum \cite{Chui71}
\begin{equation}\label{eq:Chui}
 \bigg|R_M[f(x)] - \int_{0}^{1}\!\!\mathrm{d}x\, f(x)\bigg|\leq \frac{T(f^{\prime})}{8M^2},
\end{equation}
which holds when $f$ is differentiable everywhere on $[0,1]$ with bounded derivative and total variation $T(f^{\prime})=\int_{0}^{1}\!\!\mathrm{d}x\, |f^{\prime}(x)|$.  To apply this to $E_0$, we write $M=N/2$.  Then
\begin{equation}
\begin{split}
 & \frac{1}{M}\sum_{k=0}^{M-1}\sin(\pi k/M)= \frac{1}{M}\sum_{k=1}^{M}\sin(\pi (k-1)/M)\\
  = & \cos\!\left(\frac{\pi}{2M}\right)\!R_M[\sin(\pi x)]-\sin\!\left(\frac{\pi}{2M}\right)\!R_M[\cos(\pi x)]\\
  = & (1+O(M^{-2}))R_M[\sin(\pi x)]- O(M^{-1})R_M[\cos(\pi x)]\\
  = & R_M[\sin(\pi x)]+ O(M^{-2})\\
  = & 2+ O(M^{-2}).
 \end{split}
\end{equation}
To get the second line, we used that $\sin(A+B)=\sin(A)\cos(B)+\cos(A)\sin(B)$ and equation (\ref{eq:Riemann}).  In the third line, we used $\cos(x)=1+O(x^2)$ and $\sin(x)=O(x)$ for sufficiently small $x$.  To get the fourth line, we used equation (\ref{eq:Chui}) to get $R_M[\cos(\pi x)]=0+O(M^{-2})$.  And we used equation (\ref{eq:Chui}) again to get the last line.

Therefore, the ground state energy is
\begin{equation}
 E_0=\frac{N}{a} -\frac{N}{2a}\left[2+O(N^{-2})\right]=O\left(\frac{1}{Na}\right),
\end{equation}
and since $1/Na=1/L$ is a constant, $E_0$ is essentially independent of the lattice spacing.

For general lattice regularizations of quantum field theory, the energy contained in the interactions between a region $A$ and the rest of the lattice scales like $(\partial A/a^{d-1})\times 1/a$, where $\partial A$ is the boundary of $A$. To see this, note that the region $A$ has approximately $\partial A/a^{d-1}$ interaction terms with the rest, and the strength of each interaction is proportional to $1/a$.  Dimensional analysis is often enough to argue this, but one can check for, e.g., massless scalar field theory in $d$ dimensions.  In that case, the interaction term between lattice sites comes from the discrete derivative:
\begin{equation}
 (\nabla_a \phi)^2(\vec{n})= \sum_{i=1}^{d}\left(\frac{\phi(\vec{n}+a\vec{e}_i)-\phi(\vec{n})}{a}\right)^2,
\end{equation}
where $\vec{e}_i$ are lattice basis vectors.  The Hamiltonian is given by
\begin{equation}
 H=\sum_{\vec{n}} \frac{a^d}{2}\left[\pi^2(\vec{n})+(\nabla_a \phi)^2(\vec{n})\right],
\end{equation}
where $\pi(\vec{n})$ is the operator canonically conjugate to $\phi(\vec{n})$.  Because $\phi(\vec{n})$ has dimensions of $(\mathrm{length})^{(1-d)/2}$, we see that the interaction terms between sites have strength $O(1/a)$.  One can also show that the energy \textit{density} of a product state in quantum field theory is infinite \cite{Valentini90}.

\section{Lowest energy pure state with fixed entanglement}
\label{sec:Lowest energy pure state with fixed entanglement}
We wish to minimise the energy, given that the entanglement entropy is fixed.  We can do this using Lagrange multipliers (analogously to how one derives the thermal state by maximising the entropy at fixed energy; see also \cite{GRW07} for a similar calculation).  Thus, we have the Lagrangian
\begin{equation}
 \begin{split}
  \mathcal{L}(\rho_{AB})=\mathrm{tr}[\rho_{AB} H] - \mu_1S(\rho_B)+\mu_2\mathrm{tr}[\rho_{AB}],
 \end{split}
\end{equation}
where $\mu_i$ are Lagrange multipliers, and we minimise this by setting $\partial_X \mathcal{L}(\rho_{AB})=0$, where
\begin{equation}
 \partial_X f(\sigma)=\lim_{\epsilon \rightarrow 0}\frac{f(\sigma+\epsilon X)-f(\sigma)}{\epsilon}.
\end{equation}
To compute the derivative of $S(\rho_B)$, we use
\begin{equation}
 \partial_Xf(\mathrm{tr}_A[\sigma])=\partial_{\mathrm{tr}_A[X]}f(\mathrm{tr}_A[\sigma]).
\end{equation}
We also use the following formula from \cite{Ruskai05}
\begin{equation}
 \partial_X\log(\sigma)=\int_{0}^{\infty}\!\mathrm{d}u\frac{1}{\sigma+u\openone}X\frac{1}{\sigma+u\openone},
\end{equation}
where $\sigma$ is a density operator.  This implies that
\begin{equation}
 \mathrm{tr}\left[\sigma\,\partial_X\!\log(\sigma)\right]=\mathrm{tr}[X].
\end{equation}
The result is that we need to find a state $\rho_{AB}$ satisfying
\begin{equation}
 \mathrm{tr}\left[X\left(H - \mu_1\openone_A\otimes\log(\rho_B) - \mu_1 +\mu_2 \right)\right]=0
\end{equation}
but not for \textit{any} $X$ because we want to ensure that we only consider pure states.  Writing $\rho_{AB}=\ket{\psi}\!\bra{\psi}$, let
\begin{equation}
 X=\ket{\phi}\!\bra{\psi}+\ket{\psi}\!\bra{\phi}.
\end{equation}
Then, writing $Q=Q^{\dagger}=H - \mu_1\openone_A\otimes\log(\rho_B) - \mu_1 +\mu_2$, we get
\begin{equation}
\bra{\psi}Q\ket{\phi}+\bra{\phi}Q\ket{\psi}=0.
\end{equation}
But this must be true for any $\ket{\phi}$.  Choosing $\ket{\phi}=Q\ket{\psi}$, we get
\begin{equation}
 \bra{\psi}Q^2\ket{\psi}+\bra{\psi}Q^2\ket{\psi}=0,
\end{equation}
which is only possible if $Q\ket{\psi}=0$.  So we have
\begin{equation}
 \left[H - \mu_1\openone_A\otimes\log(\rho_B) - \mu_1 +\mu_2\right]\ket{\psi}=0.
\end{equation}
Note that $\mathrm{tr}_{A}\left[\ket{\psi}\!\bra{\psi}\right]=\rho_B$, so this is unfortunately not linear.

\section{Numerical method for sampling low energy and low entanglement states}
\label{app:tn}
We now briefly describe a method for sampling states which, with respect to a given Hamiltonian and bipartition, have low energy \emph{and} low entanglement. The idea is to start in a random state, and then repeatedly attempt to lower both the energy and entanglement of this state in turn. We represent the state numerically in the form of a Matrix Product State (MPS), and utilise tensor network techniques to lower the energy and entropy.

To lower the energy of the state we perform imaginary time evolution. Specifically we apply an approximation of $e^{-\tau H}$ for some $\tau>0$, and then renormalise the state, and trim down the bond dimension. We approximate $e^{-\tau H}$ by using a Suzuki-Trotter expansion~\cite{HatanoSuzuki2005}, in method similar to that used in time-evolving block decimation~\cite{BridgemanChubb2016,Orus2014,Vidal2003,Vidal2004}.

To lower the entropy we leverage normal forms of MPS. By performing successive singular value decompositions, the Schmidt decomposition of a matrix product state
\begin{align}
	\ket{\psi}=\sum_{\alpha}\lambda_\alpha \ket{l_\alpha}\otimes \ket{r_\alpha}
\end{align}
can be efficiently calculated~\cite{PerezGarciaVerstreteWolfCirac2006}. To lower the entropy we want to `sharpen' the Schmidt spectrum, by raising it to a power
\begin{align}
	\ket{\psi'}\propto \sum_{\alpha}\lambda_\alpha^{1+\epsilon} \ket{l_\alpha}\otimes \ket{r_\alpha},
\end{align}
for some constant $\epsilon$. 

By performing the above procedures successively with random parameters on random states we can sample from states in a manner that is biased towards those with low energy and low entanglement, allowing us to approximate the Pareto front of the two variables as in Figure~\ref{fig:pareto}.

The models we chose to look at are the Heisenberg anti-ferromagnet and the transverse-field Ising model both at criticality:
\begin{equation}
 \begin{split}
  H_{\mathrm{HAF}} & =\sum_{n=1}^{N-1}\left(\sigma^x_n\sigma^x_{n+1}+\sigma^y_n\sigma^y_{n+1}+\sigma^z_n\sigma^z_{n+1}
  \right),\\
  H_{\mathrm{TFI}} & =-\sum_{n=1}^{N-1}\sigma^z_n\sigma^z_{n+1}-\sum_{n=1}^N\sigma^x_n,
 \end{split}
\end{equation}
where $\sigma_n^x$, $\sigma_n^y$ and $\sigma_n^z$ are the Pauli $x$, $y$ and $z$ matrices.

\section{Entanglement temperature near the ground state in local spin systems}
\label{app:quadratic}

In Figure~\ref{fig:pareto} we see that $T^A_{\mathrm{ent}}\propto\Delta S$ close to the ground state for the Heisenberg anti-ferromagnet and transverse-field Ising model. This behaviour is in fact generic for quantum spin systems when sufficiently close to the ground state.

Consider starting with the Schmidt decomposition of the ground state \mbox{$\ket{\Gamma}=\sum_{\alpha}\lambda_\alpha\ket{l_\alpha}\otimes \ket{r_\alpha}$}, and perturbing the highest Schmidt weight
\begin{align}
	\ket{\Gamma'}\propto \sum_{\alpha}\sqrt{\lambda_\alpha^2+\epsilon \delta_{\alpha,0}}\ket{l_\alpha}\otimes \ket{r_\alpha},
\end{align}
for some $0\leq \epsilon\ll \lambda_0$. Taking the Taylor expansions, we find that 
\begin{align}
	\Delta S&=
	-\left[S+\log\lambda_0^2\right]\epsilon+\mathcal O(\epsilon^2),\\
	\Delta E&=\left[\frac{\bra{l_0r_0}H\ket{l_0r_0}}{4\lambda_0^2}\right]\epsilon^2+\mathcal O(\epsilon^3),
\end{align}
which indeed implies $T^A_{\mathrm{ent}}\propto \Delta S$ close to the ground state, as observed.

It is worth mentioning that we are primarily interested in the regime of large (but finite) $\Delta S$ extraction, as opposed to $\Delta S \ll 1$ where $T^A_{\mathrm{ent}}\propto \Delta S \simeq 0$.  For larger $\Delta S$, we expect that generically $T^A_{\mathrm{ent}}$ is far from zero.  In contrast, the small $\Delta S$ regime is analogous to thermodynamics close to absolute zero where the heat capacity vanishes.
\end{document}

%% file: Toy.tikz
%
%
\definecolor{mycolor1}{rgb}{0.00000,0.44700,0.74100}%
\begin{tikzpicture}[%
scale=\figscale
]

\begin{axis}[%
width=0.951\figwidth,
height=\figheight,
at={(0\figwidth,0\figheight)},
scale only axis,
xmin=0,
xmax=4,
xlabel={$\Delta S_0$},
ymin=0,
ymax=2,
ylabel={$\Delta E$},
axis background/.style={fill=white},
title style={font=\bfseries},
title={Toy Model (n=5)},
legend style={legend cell align=left,align=left,draw=white!15!black},
ylabel near ticks,
xlabel near ticks
]
\addplot [color=mycolor1,solid,mark=o,mark options={solid},forget plot]
  table[row sep=crcr]{%
4	2.00000000078074\\
3.41503749927884	1.79289326554201\\
3	1.50000004421629\\
2.67807190511264	1.43211066132566\\
2.41503749927884	1.29289429875596\\
2.1926450779424	1.17712456092437\\
2	1.00000003963558\\
1.83007499855769	0.977264615964584\\
1.67807190511264	0.932113297070759\\
1.5405683813627	0.882862319161903\\
1.41503749927884	0.792893857044096\\
1.29956028185891	0.750858308800551\\
1.1926450779424	0.67713492634777\\
1.09310940439148	0.607611397151917\\
1	0.500000002401108\\
0.912537158749661	0.491666861708154\\
0.830074998557688	0.4772746310833\\
0.752072486556415	0.461087982585845\\
0.678071905112637	0.432128993181619\\
0.60768257722124	0.41413236979409\\
0.540568381362703	0.382862159598559\\
0.476438043942987	0.349367154347279\\
0.415037499278844	0.292895660359464\\
0.356143810225276	0.27711041868519\\
0.299560281858908	0.250879225470743\\
0.245112497836532	0.222977805329986\\
0.192645077942396	0.177140594566878\\
0.142019004872428	0.150453528142894\\
0.0931094043914813	0.107615685166089\\
0.0458036896131251	0.0649579039561679\\
0	-5.77315972805081e-15\\
};
\addplot [color=black,dashed,forget plot]
  table[row sep=crcr]{%
0	0\\
4	2\\
};
\end{axis}
\end{tikzpicture}%

%% file: HAF.tikz
%
%
\definecolor{mycolor1}{rgb}{0.00000,0.44700,0.74100}%
\definecolor{mycolor2}{rgb}{0.85000,0.32500,0.09800}%
\definecolor{mycolor3}{rgb}{0.92900,0.69400,0.12500}%
\definecolor{mycolor4}{rgb}{0.49400,0.18400,0.55600}%
\begin{tikzpicture}[%
scale=\figscale
]

\begin{axis}[%
width=0.951\figwidth,
height=\figheight,
at={(0\figwidth,0\figheight)},
scale only axis,
xmin=0,
xmax=1.2,
xlabel={$\Delta S$},
ymin=0,
ymax=0.9,
ylabel={$T_{\mathrm{ent}}$},
axis background/.style={fill=white},
title style={font=\bfseries},
title={Heisenberg anti-ferromagnet},
axis x line*=bottom,
axis y line*=left,
legend style={at={(0.03,0.97)},anchor=north west,legend cell align=left,align=left,draw=white!15!black},
ylabel near ticks,
xlabel near ticks
]
\addplot [color=mycolor1,solid,line width=2.0pt]
  table[row sep=crcr]{%
0	0\\
0.000226027450021982	0.000147545700164821\\
0.00114507864144175	0.000747613559614435\\
0.00232670998016393	0.00151943752865331\\
0.00457530499340564	0.00298916363030013\\
0.00695786517163344	0.00454786533145725\\
0.0108206908552678	0.00707811479179548\\
0.0148612416409769	0.00972901276859169\\
0.0194938175668579	0.0127737726835883\\
0.024272781969056	0.0159209969225536\\
0.0300497150703689	0.0197341486255048\\
0.0351354082295263	0.0230991463959902\\
0.0414662334345123	0.027298931560144\\
0.0487465536931488	0.0321440839042995\\
0.0565010927196417	0.0373237343562391\\
0.0642858075613703	0.0425439244377057\\
0.0718151150683197	0.0476130433068416\\
0.0799969045367226	0.0531448616060944\\
0.0879355447645302	0.0585365227147118\\
0.0975035731631797	0.0650679011046288\\
0.106579507426676	0.0712982692438334\\
0.11693144105486	0.0784480520182013\\
0.127003866978867	0.0854514201965591\\
0.137642273399668	0.0929008044199398\\
0.150534923875871	0.102004944741409\\
0.162443998400359	0.110493084063066\\
0.174117269488883	0.118890736490385\\
0.185546265877947	0.12719133743353\\
0.196677444777854	0.135354674985388\\
0.208562775690585	0.144162077903744\\
0.220980823049831	0.153470533997921\\
0.234355278448737	0.163625084532714\\
0.248989228041601	0.174899949919716\\
0.258435750737733	0.182274997367211\\
0.27239119948549	0.193318538320406\\
0.284769842696559	0.203271678535379\\
0.297784630006554	0.21390719160396\\
0.309814969954389	0.223904704845166\\
0.321615965201616	0.233877729125335\\
0.334033852255684	0.244562043764298\\
0.345999773386725	0.255054853185291\\
0.358777816730044	0.266489701430002\\
0.37068142430874	0.277371887010203\\
0.383280883866588	0.289151013274907\\
0.395694363803586	0.301039991471878\\
0.405693692478518	0.310837960629367\\
0.416506175123847	0.321672151938698\\
0.426055939279997	0.33146330732828\\
0.435927525307174	0.341820217604791\\
0.447471483343784	0.354260531259871\\
0.458296670454932	0.366276808402816\\
0.469862404062378	0.379526606625054\\
0.478777666317691	0.390057416822385\\
0.488880306011448	0.402356975557389\\
0.498125336718018	0.413985258443674\\
0.507795642668953	0.426568316601187\\
0.516112499223804	0.437767975973416\\
0.523333596630827	0.447802874606874\\
0.530332769838572	0.457830225935782\\
0.537576630313414	0.468548508927887\\
0.544607176916464	0.479313877568469\\
0.552697976372534	0.492192171503399\\
0.560032849237134	0.504371631211992\\
0.567224191265999	0.516835811200227\\
0.573450340854351	0.52809590359504\\
0.579529081907146	0.539560148403133\\
0.585359131953544	0.551045293779383\\
0.590996789773041	0.562665177564772\\
0.595725807423183	0.572850432722089\\
0.600932211139184	0.584586181567561\\
0.605458942839571	0.595292575852619\\
0.609557923313338	0.605444752765214\\
0.613345078307604	0.615261241609986\\
0.617250282626859	0.625881705506455\\
0.621136843975422	0.637025562267892\\
0.624027739658012	0.645739271250073\\
0.626662178937529	0.654038638098849\\
0.629539488470374	0.663545902011934\\
0.631738414744196	0.671164378468961\\
0.634165342011295	0.679977172915748\\
0.636639913154651	0.689463401743605\\
0.638256295987188	0.695972007128397\\
0.639953498822326	0.703108648270928\\
0.641746016439872	0.711028068149456\\
0.643392306719442	0.718698547042373\\
0.645095987041463	0.727102820027929\\
0.646330062085664	0.733534689444054\\
0.647521863392762	0.740064724877562\\
0.648726263971228	0.74703143600621\\
0.649936314102933	0.754467603109305\\
0.650912998264254	0.760847027499493\\
0.651612054759189	0.765655348100588\\
0.652318118208844	0.770748962160381\\
0.652963429619691	0.775643445310937\\
0.653595260606633	0.780691841488235\\
0.654177257718449	0.785603076389261\\
0.654752964613225	0.790750964002203\\
0.65522125830296	0.795188940772986\\
0.655612113950621	0.7990953425848\\
0.655992812270749	0.803107109310219\\
0.656371013532402	0.807330306235079\\
};
\addlegendentry{8};

\addplot [color=mycolor2,solid,line width=2.0pt]
  table[row sep=crcr]{%
0	0\\
0.000456258371160034	0.000230521086431929\\
0.00139746545460673	0.000706290811763351\\
0.00299482723930145	0.00151408227807863\\
0.0052599904100622	0.00265978334895584\\
0.00773163309800251	0.00391013845092162\\
0.0107278474751151	0.00542616935722669\\
0.0145888219425119	0.00738026623497657\\
0.0193994794301708	0.00981586572758402\\
0.0263278157439241	0.0133254207153551\\
0.0332289554066679	0.0168234820774441\\
0.0398258218284571	0.0201696323470715\\
0.0456826936891842	0.0231424827046892\\
0.0533815308078535	0.0270534448750271\\
0.0592965247064651	0.0300608376885191\\
0.0670342762596523	0.0339986658725246\\
0.0753203130088034	0.0382204834513955\\
0.0835300148095948	0.0424088532564815\\
0.0933489837139242	0.0474259074292538\\
0.104994584171007	0.0533880511479139\\
0.115501498172573	0.0587791287978642\\
0.126793065955577	0.0645864820444614\\
0.137217400429251	0.0699613768663203\\
0.147549248207316	0.0753024049979768\\
0.159630784501658	0.081566678493056\\
0.172362083645895	0.0881914027975204\\
0.18554470743351	0.0950784804690895\\
0.201190703003705	0.103291913745995\\
0.216598914072469	0.111425973478145\\
0.233369812458611	0.120335367058363\\
0.249449580609806	0.128937302160597\\
0.262875226961547	0.136167920196678\\
0.276010335108515	0.143288011790466\\
0.29383321579982	0.153027479656917\\
0.311360601491532	0.1627007507988\\
0.32636559449076	0.171063432184182\\
0.342938019268227	0.180394308708227\\
0.357784188485476	0.18884415039548\\
0.373419821215258	0.197843533838057\\
0.386305221245805	0.205342840754857\\
0.403886446422703	0.215705282433805\\
0.420541682685974	0.225671594194862\\
0.434461922093091	0.234122212168642\\
0.446977128661544	0.241820821649975\\
0.461196740049773	0.250692105245381\\
0.474336364657242	0.259015371370164\\
0.489635461372261	0.268870479514583\\
0.5047381802927	0.278785982020612\\
0.517967940180649	0.28763693064137\\
0.530475048830378	0.296157358535329\\
0.548207551625484	0.30851443488634\\
0.560146982286354	0.317033252904655\\
0.573605001001701	0.326844763198284\\
0.587916013115522	0.337542506579357\\
0.599606973729039	0.34650200412424\\
0.613161965149396	0.3571611409048\\
0.624127434746904	0.366015866265389\\
0.635734336567417	0.375635305184131\\
0.645956154220044	0.384335237229048\\
0.655761987345842	0.392899690436667\\
0.666639460287168	0.402672850583949\\
0.675862405054839	0.411204436059052\\
0.687750957630533	0.422567258964485\\
0.69664433570436	0.431363454105293\\
0.707237564865406	0.442208385682654\\
0.717318463191267	0.452941120415095\\
0.726454937623473	0.463057149130285\\
0.734011538009634	0.47173493966241\\
0.741057213767211	0.480106689874134\\
0.748042996047209	0.488702159926113\\
0.755370571173804	0.498067838799026\\
0.761521468339806	0.506236353593057\\
0.767611530243605	0.514630849005492\\
0.77259123674065	0.521745373692035\\
0.777505907343379	0.529010713570077\\
0.782800146401725	0.537136296472561\\
0.787450080100936	0.544556677010557\\
0.791022859192764	0.550457173628294\\
0.795427461319259	0.557994294018088\\
0.799579691929324	0.565393772188604\\
0.803558458515622	0.572783733367582\\
0.80788890324199	0.58120257657677\\
0.811086475253911	0.587702073682683\\
0.81447789293699	0.594892409299532\\
0.817487367722899	0.601560276695375\\
0.82009712007348	0.607588345518677\\
0.823209688659635	0.615115476669883\\
0.82627959728086	0.622951218273635\\
0.829185227765527	0.630806429368731\\
0.831610457370645	0.637744565121446\\
0.833512505583494	0.643467499323391\\
0.835168938042596	0.648681891455066\\
0.836628321245364	0.653476927310199\\
0.838447015035626	0.659753120279368\\
0.839852663759437	0.664865840615549\\
0.841452910383718	0.671010819984341\\
0.842585141994904	0.67560020364064\\
0.843640386019731	0.68008636617033\\
0.84454403539257	0.684110902766637\\
0.845645731684975	0.689281248693852\\
};
\addlegendentry{16};

\addplot [color=mycolor3,solid,line width=2.0pt]
  table[row sep=crcr]{%
2.92355454030169e-05	-9.81617074660535e-06\\
0.000354322014876507	0.000123929563482218\\
0.00142172971079035	0.000570167298598511\\
0.00332416195353136	0.0013656420805241\\
0.00590869678703632	0.00244617688803375\\
0.00828996150368377	0.00344155288384724\\
0.0121674470135167	0.00506199216854877\\
0.0166405945591181	0.00693081962739934\\
0.0217057453735388	0.00904629707596229\\
0.0271180447873729	0.0113059936405185\\
0.0334494228569877	0.0139484564471301\\
0.0403557841183384	0.0168297822264204\\
0.0490309977627792	0.0204475489028943\\
0.0582816520028102	0.0243035953851668\\
0.068333118094049	0.0284917118327005\\
0.0781917444945249	0.0325979792538055\\
0.0896410819192083	0.0373653032756958\\
0.101234023278486	0.0421912168421525\\
0.112678821529554	0.0469547421073884\\
0.123743210387094	0.0515597122689844\\
0.135593207777452	0.0564919315911048\\
0.147058783788158	0.0612649590610839\\
0.158373567270214	0.0659765370737278\\
0.171069989150079	0.07126568967531\\
0.18677562036488	0.0778128904782742\\
0.20028305746497	0.0834488050338915\\
0.213139259860295	0.0888183755048944\\
0.227978967320552	0.0950241788219541\\
0.241726054344901	0.100781788926983\\
0.256318046925476	0.106903912736509\\
0.275139087253809	0.114819091363925\\
0.290284666809484	0.121206147172069\\
0.305705166064275	0.127727471367568\\
0.321765058150724	0.134541223133299\\
0.337486161557969	0.141235497733398\\
0.357276280512941	0.149700593536991\\
0.37287569835602	0.156406471338293\\
0.392322784202659	0.164812233954458\\
0.410044871714837	0.172521369151873\\
0.426706226003466	0.179816097920866\\
0.443994288709463	0.187438025435993\\
0.459883798690256	0.19449509484869\\
0.476997288626676	0.20215617002557\\
0.495328381560001	0.210438065991224\\
0.511866975418549	0.217983408992664\\
0.527915181154375	0.225377288553165\\
0.543963408529466	0.232848253244233\\
0.560055902467374	0.240423558834909\\
0.576654316264369	0.248332150321342\\
0.591812821604136	0.255645958924961\\
0.60896037664973	0.264033346229341\\
0.626688658618801	0.272842957518855\\
0.63984060005065	0.279476878458605\\
0.655527605978552	0.287508195235602\\
0.676441611033049	0.298434680405688\\
0.691056572500519	0.306233027110671\\
0.706861219762069	0.314831645435952\\
0.7203723995582	0.322330495469852\\
0.735051285328217	0.330645707779804\\
0.749640170493801	0.339098945917451\\
0.763037635234399	0.347042938045946\\
0.77516796255132	0.354398374598439\\
0.789653918296707	0.363403478523303\\
0.801846305894905	0.371186731018402\\
0.814674889329709	0.379597080476094\\
0.826279066803081	0.387418388587962\\
0.837089361275792	0.394904585604653\\
0.847130643942744	0.40204681176413\\
0.858319175274469	0.410239226378932\\
0.867403431711421	0.417089392171432\\
0.87787516168337	0.425228619922801\\
0.888339738686182	0.433648834936026\\
0.897200218071258	0.441026433224682\\
0.9062090910589	0.448786175806054\\
0.91469432558645	0.456358849407589\\
0.923572970220972	0.46458805267021\\
0.929833652330725	0.470598826240926\\
0.935928945590721	0.476633044432228\\
0.942127966066076	0.482971817560987\\
0.94828222808143	0.489486274830582\\
0.953857002619827	0.495597342076103\\
0.959199392920867	0.501660424816872\\
0.965044326013799	0.508550867494595\\
0.969809851620071	0.51438947576835\\
0.974750059920997	0.520675365078548\\
0.978581036482029	0.525730710278109\\
0.98249089190935	0.531070271498006\\
0.986530359838968	0.536798049135607\\
0.990526615037176	0.54269992139388\\
0.994395198429462	0.548663601790477\\
0.998316847410472	0.554993580471537\\
1.00183534011721	0.560950034298056\\
1.00483605064509	0.566265502473011\\
1.00759156182799	0.571362827059644\\
1.01018593694742	0.576375606498811\\
1.01272975415281	0.58151835443365\\
1.01448784692608	0.585220674103175\\
1.01652727672064	0.589685722570851\\
1.01855753980331	0.594334678118843\\
1.02045120854773	0.598879374417721\\
};
\addlegendentry{32};

\addplot [color=mycolor4,solid,line width=2.0pt]
  table[row sep=crcr]{%
0	0\\
0.00234333825672417	0.000844614610245379\\
0.00422913424412319	0.00145308389854618\\
0.00657582403532508	0.00226794431112669\\
0.0103184994686756	0.00359797137835501\\
0.0141507222090309	0.00497075298145184\\
0.0190156205376928	0.00671831658797422\\
0.0258321627365987	0.00916945310987686\\
0.0319969080366063	0.011386281791492\\
0.0395898367343472	0.0141154670485602\\
0.0459111209838066	0.016386125392427\\
0.0546457904451612	0.0195212529831182\\
0.0628203485409213	0.0224524311907462\\
0.0728576486279302	0.026047834618968\\
0.0811784889605724	0.0290252080212466\\
0.0912300657326308	0.032618059934908\\
0.101093339025065	0.0361395944281223\\
0.111960913789898	0.0400151848369999\\
0.121750274985183	0.0435023244698377\\
0.135369958303879	0.0483478791008533\\
0.147869629950603	0.0527890442513682\\
0.162017136255497	0.0578092100005751\\
0.177489389925119	0.0632920635157116\\
0.190497472824716	0.0678962121152034\\
0.201733862537624	0.0718693916744895\\
0.21605763022015	0.0769297205188873\\
0.233381810312324	0.0830438874924283\\
0.249803061931957	0.0888340988616903\\
0.266015610760123	0.0945466515228296\\
0.284080015798375	0.10090820952256\\
0.300732695244632	0.106770574398981\\
0.31820826735345	0.112921935882544\\
0.336606331504563	0.119398890169206\\
0.353063460753518	0.12519483197777\\
0.368076613950108	0.130485368262321\\
0.386617938773584	0.137025103025456\\
0.404413328087001	0.143309839737675\\
0.422948240042015	0.149866527794417\\
0.440353174992714	0.156035746605037\\
0.459111524274507	0.162700613220835\\
0.478285533326549	0.169533268318721\\
0.494593075891	0.17536296470988\\
0.5120056622369	0.181609110591306\\
0.530229250751164	0.188172897134903\\
0.548262263604528	0.194698307364992\\
0.567454395263809	0.201680271079969\\
0.586103261715639	0.208505574334681\\
0.606890041815935	0.216166551838864\\
0.62326095341968	0.222243845083814\\
0.645242003708493	0.23047123064409\\
0.660227552741995	0.236128916769318\\
0.67503307506866	0.241761002076176\\
0.694229291797086	0.249131482878503\\
0.709832692673275	0.255184134336347\\
0.731547870081007	0.263708749174258\\
0.750592304643689	0.271291429134211\\
0.770176949067079	0.279204036525513\\
0.78875830009622	0.28682984241638\\
0.801959882347103	0.292324307882627\\
0.819379518188193	0.299680197318126\\
0.836854318513364	0.307191047703942\\
0.852490543184157	0.314033945551633\\
0.865066459874734	0.319628576570976\\
0.881449731471907	0.327049183139824\\
0.894947591338764	0.333284697275748\\
0.910355558211589	0.340549529980547\\
0.924592288065408	0.347413940214261\\
0.936484060675211	0.353269475594814\\
0.950341904596401	0.360245442654578\\
0.963601058035338	0.367087377532138\\
0.975220014954378	0.373229962823259\\
0.986627388549031	0.379406369137497\\
0.997556929621939	0.385471330784816\\
1.010202035459	0.392685154987209\\
1.02096536507666	0.399008007473907\\
1.03049820721332	0.404761841053335\\
1.03965186648172	0.410435156420866\\
1.04885923807812	0.416301596161586\\
1.05956279636935	0.423342798532344\\
1.06910189338561	0.429839927932168\\
1.07876636241765	0.43666016548553\\
1.08725439836389	0.442869697094274\\
1.09413459868009	0.448069758889741\\
1.10215223934316	0.454337917615213\\
1.11066582164981	0.461267370956014\\
1.11688613822371	0.466529360251654\\
1.12523404471039	0.473889394875453\\
1.13039851447973	0.478633916468008\\
1.13581381433305	0.483784703482567\\
1.13993903942096	0.487841107880207\\
1.14492750188015	0.492915604843613\\
1.14942142547102	0.497662122686278\\
1.15348370965854	0.502111253078196\\
1.15807725781147	0.507344375038862\\
1.16154593024024	0.511454214014506\\
1.1660098396374	0.516968882809791\\
1.16951323206216	0.521496889054204\\
1.17277271353527	0.525888318624038\\
1.17643538707466	0.531055834812599\\
};
\addlegendentry{64};

\end{axis}
\end{tikzpicture}%

%% file: TIM.tikz
%
%
\definecolor{mycolor1}{rgb}{0.00000,0.44700,0.74100}%
\definecolor{mycolor2}{rgb}{0.85000,0.32500,0.09800}%
\definecolor{mycolor3}{rgb}{0.92900,0.69400,0.12500}%
\definecolor{mycolor4}{rgb}{0.49400,0.18400,0.55600}%
\begin{tikzpicture}[%
scale=\figscale
]

\begin{axis}[%
width=0.951\figwidth,
height=\figheight,
at={(0\figwidth,0\figheight)},
scale only axis,
xmin=0,
xmax=0.8,
xlabel={$\Delta S$},
ymin=0,
ymax=0.375,
axis background/.style={fill=white},
title style={font=\bfseries},
title={Transverse-field Ising},
axis x line*=bottom,
axis y line*=left,
legend style={at={(0.03,0.97)},anchor=north west,legend cell align=left,align=left,draw=white!15!black},
ylabel near ticks,
xlabel near ticks
]
\addplot [color=mycolor1,solid,line width=2.0pt]
  table[row sep=crcr]{%
0.00435760871385005	0.00350747505218108\\
0.00786372205468244	0.006164669252637\\
0.0124930596514604	0.0098117406155733\\
0.0153828224686706	0.012151609407853\\
0.019041810053514	0.0149434483653966\\
0.0215449419044521	0.0169315793412439\\
0.024613795579281	0.0193905577611306\\
0.0270440215324225	0.0211707724534122\\
0.0297146704020773	0.0231712858099034\\
0.0322483437379554	0.0251789644057511\\
0.0351029737673675	0.0275053161205189\\
0.0376378108369353	0.0294704456022119\\
0.0404721999135698	0.0315501327848629\\
0.0432359425852342	0.033713053792576\\
0.0458548863849114	0.035699699072414\\
0.048328069235321	0.03760258450836\\
0.0506916234151428	0.0395850638621552\\
0.0538804076853877	0.0418683304287706\\
0.0565268994300231	0.0439450527681719\\
0.0590247736428006	0.0459033048502258\\
0.0609075762252403	0.0473384757324118\\
0.0630629742020774	0.0489867372278724\\
0.0650961432258164	0.050605622517227\\
0.0673284289314425	0.052259048967643\\
0.0690980300106555	0.053655343719772\\
0.0720794886609825	0.0559212832453448\\
0.0740980564782942	0.0576331888959759\\
0.0761643808583866	0.0590549620476456\\
0.0783501618146796	0.0608162868210021\\
0.0805420682490014	0.0625356325216906\\
0.0826788039423981	0.0640756846308174\\
0.08454561345358	0.0655514083132576\\
0.0868712067050526	0.0673136417824914\\
0.0891499220548236	0.0690877230849888\\
0.091125152159843	0.0706244330460324\\
0.0927999386980805	0.0718905179868448\\
0.0945203221167389	0.0731816527029789\\
0.0963143056937046	0.0745802870214405\\
0.0983873540357216	0.0762454527935854\\
0.100521797942942	0.0778275809114422\\
0.10213823002636	0.0790853694956147\\
0.103733635164209	0.08040275768768\\
0.105366445207645	0.0815945932756355\\
0.107439581054244	0.0831839201895132\\
0.109230643788694	0.0845115109480094\\
0.110902903275184	0.0858351316538254\\
0.112484161041898	0.0870252759831573\\
0.113991099381904	0.0882207213620633\\
0.115759661396558	0.0895899854769831\\
0.117771753661683	0.091120734273354\\
0.11946683625782	0.0924500411873032\\
0.121355176535113	0.093844667484212\\
0.122724101223476	0.0949247561769061\\
0.124480415738465	0.0962869866277724\\
0.125807377719943	0.0972911775483032\\
0.127366371317411	0.0985087591147408\\
0.129011593671713	0.0997726649638957\\
0.130481028649921	0.100892405899939\\
0.131884322109561	0.102008624473259\\
0.133348160129802	0.10312342664867\\
0.134656985661303	0.104113318980543\\
0.135815002344349	0.105008836170225\\
0.137052372658985	0.106019586492302\\
0.138218489155009	0.106925455075972\\
0.139589653449882	0.107969693074779\\
0.140838939439239	0.108979384241701\\
0.142305667643542	0.110061857596786\\
0.143723818105829	0.111138354406762\\
0.14506427466648	0.112171015479207\\
0.146274582674507	0.113144546738372\\
0.147294864078883	0.113886843329853\\
0.148374653585498	0.114729966334884\\
0.149421428397182	0.115585221046622\\
0.150492824313482	0.116398355287233\\
0.151394676242478	0.117146699495176\\
0.152158693286405	0.117664309379111\\
0.153213810883702	0.118516861080525\\
0.154110502955668	0.119184779405648\\
0.154925883022037	0.119808315561029\\
0.155630918856687	0.120377193624217\\
0.156451639794693	0.120957813568754\\
0.157259209866335	0.121569453576935\\
0.158391536252387	0.122408384798231\\
0.15951085149526	0.123236144229066\\
0.16113591252415	0.124478811543702\\
0.163226078440011	0.125874663221733\\
0.166183173931554	0.128000451023937\\
0.169591181283165	0.130232288685669\\
0.174092297050222	0.133319472957285\\
0.178945380440261	0.136390021748301\\
0.186872459768307	0.140729148902685\\
0.194179347546142	0.144760960283298\\
0.205968650353135	0.151141474783148\\
0.225786757396272	0.160614073004749\\
0.245415068780046	0.168455531074607\\
0.290015855211717	0.184496635973274\\
0.343418062102859	0.203706436624822\\
0.412500860028078	0.229808091080679\\
0.476506980037335	0.264739840059927\\
0.509860043276131	0.310234829158277\\
};
\addlegendentry{8};

\addplot [color=mycolor2,solid,line width=2.0pt]
  table[row sep=crcr]{%
0.00315777267910611	0.00250230716618368\\
0.00677469295557698	0.00485501354240993\\
0.00907522474855016	0.00618521470023077\\
0.0112257804808636	0.00766361862215414\\
0.0137507642262159	0.0091998319657198\\
0.0163507302160989	0.0108073853711608\\
0.0186619646545302	0.0122570917651194\\
0.0202109221644803	0.013450363784161\\
0.0231855221668158	0.0152522389299694\\
0.02430741587187	0.016152876337175\\
0.0261002482952005	0.0170996393511747\\
0.0279656645132134	0.0185301683744546\\
0.0296615301928262	0.0196410767679779\\
0.0323519316189853	0.0214353743365294\\
0.0346962557847709	0.0229470291848129\\
0.0362936694016942	0.0239730163136346\\
0.0385008171388939	0.0254277112566205\\
0.0407454865280057	0.0269319384867616\\
0.0421491805646058	0.0278235228240594\\
0.0439768763728737	0.0290982977504462\\
0.0468459117449999	0.0308552222363036\\
0.0491087959561253	0.0321647265089322\\
0.0503674900342908	0.033330870916621\\
0.0516910229586118	0.0340430146905655\\
0.0533388530251901	0.0352261340198009\\
0.0552578853262869	0.0363101427983158\\
0.0564680103311347	0.037010301109448\\
0.0583980936448584	0.0383055702763747\\
0.0603355044220294	0.039514070287294\\
0.0619572325243966	0.0406279409719238\\
0.063133198588154	0.0414425712623137\\
0.0645448211522434	0.0423691186067824\\
0.0659445519507588	0.0433481367069958\\
0.0679986204628276	0.0448326634794884\\
0.0693184674291516	0.045473434586841\\
0.0709363866671948	0.0463924892196323\\
0.0727784665660378	0.0477027648549837\\
0.0734188274281818	0.0481843848444021\\
0.074861706045363	0.0490625308739425\\
0.0762610889015088	0.0500098379036129\\
0.0781674064089066	0.0511769765701972\\
0.0794870384605872	0.0520288851885303\\
0.0808910111623811	0.0528511453147245\\
0.0821576384553465	0.0537343897242305\\
0.0840544203256516	0.0549830448571747\\
0.0854609671228713	0.0559011209234199\\
0.0863810021909051	0.0564405510890408\\
0.0877774200836333	0.0572464772380268\\
0.0893601851653829	0.0581920727184857\\
0.090317899823772	0.0586973479992068\\
0.0915783645620737	0.0595171679215191\\
0.0930280932629395	0.0602815292876462\\
0.0948582222149973	0.0612448720794101\\
0.0967283068134442	0.0621564025881155\\
0.0986363638384216	0.0628139417988834\\
0.101322186998932	0.06418174037022\\
0.1045014005369	0.0657081601825066\\
0.106552487295291	0.0664926542694624\\
0.10861095105367	0.0672688671124745\\
0.111992210325834	0.0686568005173541\\
0.114428084677184	0.0696830966248415\\
0.118606253176488	0.0709654533422746\\
0.122896047496468	0.0725785782915607\\
0.127249282618768	0.0737467975071929\\
0.132433067027034	0.0752917158634183\\
0.135330988141855	0.0761878048769509\\
0.141305238674004	0.0777462023137922\\
0.149322140243518	0.0795822224131213\\
0.154152219630759	0.080678341835299\\
0.162436951617873	0.082454425236789\\
0.1692715241916	0.0839359157761959\\
0.178814426302811	0.0857101835368237\\
0.187659174433398	0.0870750885029004\\
0.197250517403738	0.0886154113470974\\
0.207693461010108	0.0905275196900605\\
0.22207443486713	0.0924577553850258\\
0.23872186306083	0.0946808282015661\\
0.25121953169911	0.0962217999695149\\
0.266927221581687	0.0980701356392936\\
0.282255138448106	0.0999097553486867\\
0.296146463624664	0.101682493925348\\
0.313043771168742	0.10344491321848\\
0.332682758204425	0.105964632403984\\
0.356179515582878	0.10847145633058\\
0.375187471434998	0.110929964626224\\
0.401325539870133	0.114380224253891\\
0.415451220604839	0.117178838655544\\
0.445445886559303	0.121600626512995\\
0.467764544637799	0.125921228975549\\
0.49313854624538	0.131233003363652\\
0.510527234549438	0.136773913345943\\
0.523348914126208	0.140323989727117\\
0.538751506571467	0.146029305494181\\
0.55890258139148	0.155806395068319\\
0.575245532802967	0.164771776289749\\
0.585735596868619	0.173788680323466\\
0.59445412476883	0.184042004016651\\
0.602270486279993	0.193904176175938\\
0.605718410846869	0.200122649456703\\
0.608155133794258	0.207248769490885\\
0.609820946420468	0.212803856776891\\
0.610722725957209	0.218881072527141\\
};
\addlegendentry{16};

\addplot [color=mycolor3,solid,line width=2.0pt]
  table[row sep=crcr]{%
0.00305374925181157	0.00254249838775132\\
0.00793870111667039	0.00375700338336681\\
0.011072250446197	0.00542609620784872\\
0.0134375637033041	0.00674702236483187\\
0.0215695702309325	0.00920903816156922\\
0.0253711927221908	0.0116660310432531\\
0.027740832058106	0.0128049337808691\\
0.0409195952938813	0.0167722252930815\\
0.0543167592211043	0.0215146784434607\\
0.0632199845103061	0.0308324898094108\\
0.0723573952740459	0.0313349048131143\\
0.0870713520585356	0.034768354883145\\
0.0952469576181975	0.0374284508515811\\
0.10427933858672	0.0385191065656161\\
0.110174785368451	0.0399790909913289\\
0.147183941260567	0.0417900975445803\\
0.178261458683126	0.0461636832938163\\
0.205259706654569	0.0481818639302066\\
0.213408912746003	0.048579806978482\\
0.227855576996576	0.0491260777217823\\
0.233831981728848	0.0499343749664421\\
0.240139844817825	0.05018140566214\\
0.265987729006244	0.0507971766817699\\
0.284716258182713	0.0523480286054662\\
0.297329356124811	0.0528601719914265\\
0.306276582529575	0.0533896600760163\\
0.317568523126026	0.0539784015610974\\
0.334101254201175	0.054726331450566\\
0.344886551483278	0.0551055225199584\\
0.351683068526869	0.055754935313291\\
0.362837909810302	0.0559812827990341\\
0.368479233804263	0.0565321741373242\\
0.377142899104634	0.0572753919227458\\
0.391359808136099	0.0578082167302429\\
0.392149716593656	0.0580315053264953\\
0.403449406246355	0.0584122658333825\\
0.413654573919901	0.0592788490273336\\
0.42703196848455	0.0599113980105473\\
0.431765791440665	0.0605349060479844\\
0.442237537002598	0.0612875837597517\\
0.450041394308848	0.0619163833663779\\
0.455995964072428	0.0627117133181128\\
0.457561121088787	0.0630461478249507\\
0.464187460619696	0.0634085410466159\\
0.469689741744097	0.0638281011169369\\
0.477630566758678	0.0644248903973562\\
0.485305443351283	0.065139885160607\\
0.4889985350108	0.0658634070585906\\
0.494211816872201	0.0662931954524691\\
0.496249937469628	0.0667776301641291\\
0.500717824983613	0.0671752615243741\\
0.508145427521334	0.0678622984520766\\
0.513458525317286	0.0686541023809643\\
0.52103607768742	0.0696439675004089\\
0.525961903502757	0.0708923481384042\\
0.532071239646715	0.0715881859707359\\
0.540853279905101	0.0724707377198034\\
0.547458036315777	0.0738462197523469\\
0.554184183654839	0.0754117883612819\\
0.558994610732192	0.0765019529739099\\
0.562659473760149	0.0768504125909523\\
0.565358650466519	0.0780974046324413\\
0.56937774446708	0.0784317111716629\\
0.574160167891977	0.0799085914271137\\
0.581016964612946	0.0806033312353998\\
0.581768536820191	0.081618021135999\\
0.588688165383105	0.0829520560739311\\
0.593593433324168	0.0841258418114894\\
0.600438468650574	0.0859732316871275\\
0.604669463794389	0.0880959194318772\\
0.608215230111867	0.0891799846545707\\
0.613210468042835	0.0911836656448885\\
0.618901157314835	0.0929888245541001\\
0.628301459942975	0.095315849061867\\
0.629357538893708	0.0977467959125093\\
0.633047527527277	0.0995401204944165\\
0.637270381435344	0.10148530957223\\
0.641726042376879	0.103042864919133\\
0.643987258089644	0.104089787796485\\
0.645474063415271	0.104775353423833\\
0.648443892924462	0.106596824298111\\
0.65209530721331	0.108779971442754\\
0.658441003168093	0.110996515183216\\
0.659740281339956	0.113454063006761\\
0.663234994507439	0.115042178770362\\
0.665645729696463	0.117121009516576\\
0.669354925636605	0.118299171454027\\
0.672237436905754	0.121679292487879\\
0.67380624732612	0.123625042625077\\
0.675649924133773	0.124090045890523\\
0.676943900914169	0.125745578727572\\
0.680157244853092	0.128802043849538\\
0.682736567960757	0.130788735457063\\
0.68436419245245	0.134545477909083\\
0.686537458078234	0.135647338496938\\
0.687166940307764	0.136605487879513\\
0.690090634265355	0.138597647316118\\
0.69127476739461	0.141995915649041\\
0.691930583901867	0.142868476755718\\
0.69270835533287	0.143435177139944\\
0.693810936943123	0.145477774522336\\
0.694419802229029	0.146976961669536\\
0.694654747085487	0.148001729896004\\
0.69507653892178	0.148735852050477\\
0.69601683924197	0.150139013936319\\
0.696339836396764	0.151357610056752\\
0.697728287696985	0.153322738472476\\
0.698221578517678	0.1546449252662\\
0.698548604898362	0.156100080687322\\
0.698949877863242	0.156469785684056\\
0.699326588118936	0.157611505992242\\
0.699472238435701	0.158785800017353\\
0.699837162513529	0.160022535482087\\
0.700030684860605	0.160980034746164\\
0.700246078903039	0.162321758509999\\
};
\addlegendentry{32};

\addplot [color=mycolor4,solid,line width=2.0pt]
  table[row sep=crcr]{%
0.0209370810529225	0.00506819306920586\\
0.0294471863309289	0.0062079000408295\\
0.0392462044491331	0.00727873365287417\\
0.0504884999961672	0.00964642178303406\\
0.0616268207333199	0.0129116284452236\\
0.0642477628627324	0.0134814794160787\\
0.0715891905118844	0.0154268434671937\\
0.0837481223593404	0.0159696408171001\\
0.0860262092083121	0.0178294489644136\\
0.0895019252961921	0.0183762185980699\\
0.0958863435025866	0.0187567562659884\\
0.107626768154205	0.0190815074044047\\
0.120232141095896	0.0199220605060471\\
0.130217493244392	0.0217462396266332\\
0.136807918402795	0.0220993785107812\\
0.151196190114279	0.0225954280836564\\
0.168878032063901	0.0229549836558426\\
0.185022779774164	0.0236428431384151\\
0.201908757058332	0.024731781790886\\
0.208224974529072	0.0251590015623102\\
0.235608301298189	0.0252912392043374\\
0.240143908315259	0.0259332945392953\\
0.251915558723228	0.0262798352689036\\
0.292648623520076	0.0268577466082108\\
0.304248670132161	0.0272989029793218\\
0.317710388083782	0.0279153473671717\\
0.321559474076611	0.0281579016048742\\
0.328855171232141	0.0285259617511255\\
0.342463440048891	0.0287188554240338\\
0.345490215433605	0.0287384785356915\\
0.350243123666631	0.0289350336759357\\
0.363248729367078	0.0291492608631212\\
0.374006383260103	0.0293812702526707\\
0.388285518383278	0.0299385777939175\\
0.417447835853686	0.0304493823868857\\
0.422679164105951	0.0309340787012423\\
0.428975903186822	0.0311416737217964\\
0.429911432114111	0.0313815117550265\\
0.432555151643735	0.0316824605122264\\
0.441216672266641	0.0318003546464578\\
0.449021497844364	0.0320845443186006\\
0.455256849327938	0.0323380913790153\\
0.463104898506198	0.0325972721509511\\
0.472070250856072	0.0329255542752174\\
0.474900162570059	0.0333403905492511\\
0.490068867184911	0.0337435298937625\\
0.495463731877433	0.0341813023956761\\
0.51085410990974	0.0348145690355691\\
0.512647101131445	0.035086528429367\\
0.516468714530664	0.0358081033877464\\
0.525411699710476	0.0360600431795156\\
0.535168359564932	0.0363678537624961\\
0.544669050819189	0.0373961484605561\\
0.550232807366893	0.0384720660851803\\
0.562760326813495	0.0388367583150194\\
0.569025440152023	0.0394732903210672\\
0.583589416677324	0.0405431062810866\\
0.592194333456491	0.0424675217466765\\
0.597577232374621	0.0427390433588698\\
0.598925851484957	0.0442184843832822\\
0.605035045389685	0.0443553877385296\\
0.620889432402971	0.0463242011482182\\
0.631791053201094	0.0496131893442675\\
0.639560656132632	0.0502945835999629\\
0.647532107599758	0.051242294205699\\
0.65399447285307	0.0534576119297886\\
0.663772982787546	0.0554492464885727\\
0.666017119400974	0.0571007390310892\\
0.669407997190783	0.0580179406736528\\
0.676941026429013	0.0603733064330583\\
0.689344470623676	0.0629783766927436\\
0.693311220394961	0.0659849258559182\\
0.699274642517868	0.0665922002198372\\
0.701173172303991	0.0686865498155988\\
0.716812242296745	0.0723587082937181\\
0.717611713181655	0.0759598668805387\\
0.720761652332964	0.0780988363788502\\
0.723429601066625	0.0790031934919221\\
0.730279690758212	0.0801625509904154\\
0.736110574837906	0.0847958622717648\\
0.741814414470663	0.087978373366803\\
0.743116528263983	0.0903143510929212\\
0.747273823658033	0.0906263305699144\\
0.750498595473639	0.0927489920814871\\
0.756059017093749	0.095940418281528\\
0.756773115828703	0.0993733014204928\\
0.760272315438339	0.102121501384641\\
0.764550513088805	0.103849346187375\\
0.768389222795177	0.108989991033405\\
0.769169809888904	0.111624788307273\\
0.769549440918407	0.112400086269141\\
0.772107183452619	0.112530243488059\\
0.772690725213406	0.113089295571506\\
0.77351980602909	0.114205901553364\\
0.774779517935244	0.116270361966522\\
0.775956214280424	0.118049507734287\\
0.777117896515461	0.120029181728656\\
0.778566862227704	0.122321817479897\\
0.780727922865235	0.124381602657045\\
0.781057266993554	0.125998148891382\\
0.781526915576647	0.127533056864202\\
0.782693449839895	0.130388675756997\\
0.783192708182853	0.131128853982494\\
0.783433578876911	0.132052872639489\\
0.78396005542399	0.133587503714788\\
0.784057168396175	0.135250947709621\\
};
\addlegendentry{64};

\end{axis}
\end{tikzpicture}%